\documentclass[showpacs,a4paper,12pt,prl,onecolumn]{revtex4}%
\usepackage{amsfonts}
\usepackage{amsmath}
\usepackage{amssymb}
\usepackage[pdftex]{graphicx}%
\usepackage{epstopdf}
\providecommand{\U}[1]{\protect\rule{.1in}{.1in}}

\ifx\pdfoutput\relax\let\pdfoutput=\undefined\fi
\newcount\msipdfoutput
\ifx\pdfoutput\undefined\else
\ifcase\pdfoutput\else
\ifx\paperwidth\undefined\else
\ifdim\paperheight=0pt\relax\else\pdfpageheight\paperheight\fi
\ifdim\paperwidth=0pt\relax\else\pdfpagewidth\paperwidth\fi
\fi\fi\fi
\begin{document}
\title{Steady Fock states via atomic reservoir}
\author{F. O. Prado$^{1}$, W. Rosado$^{2}$, G. D. de Moraes Neto$^{2}$, and M. H. Y.
Moussa$^{2}$}
\affiliation{$^{1}$Universidade Federal do ABC, Santo Andr\'{e}, S\~{a}o Paulo, Brazil}
\affiliation{$^{2}$Instituto de F\'{\i}sica de S\~{a}o Carlos, Universidade de S\~{a}o
Paulo, S\~{a}o Carlos, S\~{a}o Paulo, Brazil}

\begin{abstract}
In this letter we present a strategy that combines the action of cavity
damping mechanisms with that of an engineered atomic reservoir to drive an
initial thermal distribution to a Fock equilibrium state. The same technique
can be used to slice probability distributions in the Fock space, thus
allowing the preparation of a variety of nonclassical equilibrium states.

\end{abstract}

\maketitle

The development of strategies to prepare nonclassical states \cite{PNS} and,
in particular, to circumvent their decoherence ---via decoherence-free
subspaces \cite{DFS}, dynamical decoupling \cite{DD}, and reservoir
engineering \cite{PCZ,RE}--- have long played a significant role in quantum
optics. On the conceptual side, the need for these states stems from their use
in the study of fundamental quantum processes, such as decoherence
\cite{Decoherence} and the quantum to classical transition \cite{QC}. On the
pragmatic side, the advent of quantum computation and communication ---which
depends strongly on successfully producing highly nonclassical states and
ensuring their long-term coherence \cite{Livro}--- has certainly put extra
pressure on researchers to implement efficient techniques of engineering and
protection of nonclassical states. The proposition of schemes that enable the
generation of nonclassical equilibrium states thus represents an ideal
approach to the current challenges. In this regard, the reservoir engineering
technique proposed in Ref. \cite{PCZ} and experimentally demonstrated in a
trapped ion system \cite{IT} signals an important step toward the
implementation of quantum information processes \cite{Livro}, a goal that has
recently mobilized practically all areas of low-energy physics. Reservoir
engineering, however, has major limitations, starting with the fact that it
prevents, for example, the generation of Fock equilibrium states (a key goal
of the present letter). Moreover, the protection of a particular state demands
the (not-always-easy) engineering of a specific interaction which the system
of interest is forced to perform with other auxiliary quantum systems.

Recently, the generation of Fock states with photon numbers $n$ up to $7$ was
reported in cavity QED, where a quantum feedback procedure is employed to
correct decoherence-induced quantum jumps \cite{Haroche}. The resulting photon
number distribution assigns a probability around $0.8$ to the generation of
number states up to $3$, falling to below $0.4$ for $n=7$. Nonequilibrium
number states up to $2$ photons have long been prepared in cavity QED
\cite{Walther}, as well as in most suitable platforms, such as ion traps
\cite{Leibfried} and, lately, in circuit QED \cite{Cleland}, where number
states up to $6$ were achieved.

In this letter we present a protocol in which the atomic beam reservoir
technique \cite{ABR} is exploited to produce high fidelity equilibrium Fock
states in cavity QED. The atomic reservoir ---built up by injecting a beam of
atoms that interact, one at a time, with the cavity mode--- prompts the
emergence of an engineered Liouvillian superoperator to govern the cavity
field dynamics, alongside that coming from the cavity loss mechanisms. We
stress, from a practical perspective, that atomic reservoirs have for some
time been used for the preparation of the cavity vacuum state \cite{RMP}.
Moreover, this has been theoretically explored, in close relation to the
reservoir engineering technique \cite{PCZ}, for the generation of an
Einstein-Podolsky-Rosen steady state comprising two squeezed modes of a high
finesse cavity \cite{BR}. Our proposal, however, is altogether different from
those in Ref. \cite{BR,PCZ}. In contrast to the assumptions that support the
method proposed in \cite{BR, PCZ}, our protected Fock state is not a steady
state of a specific engineered Lindbladian $\mathcal{L}\rho=\left(
\Gamma/2\right)  \left(  2\mathcal{O}\rho\mathcal{O}^{\dagger}-\mathcal{O}%
^{\dagger}\mathcal{O}\rho-\rho\mathcal{O}^{\dagger}\mathcal{O}\right)  $,
where the only pure steady state of the system is the eigenstate of operator
$\mathcal{O}$ with a null eigenvalue. In our protocol, the steady state is
driven by a sum of three engineered Lindbladians, two of which act on selected
subspaces of the cavity mode space, the mode emitting or absorbing photons
within these subspaces. The third Lindbladian is associated with
(non-selective) photon absorption by the cavity mode, in order to
counterbalance the inevitable emission to the natural (nonengineered)
environment. The selective Lindbladians are built up from engineered selective
Jaynes-Cummings (JC) Hamiltonians, while the Lindbladian for photon absorption
follows from a usual JC interaction.

To generate the required Lindbladians, we rely on engineered selective
Hamiltonians \cite{WilsonJOPB} and engineered atomic reservoirs \cite{ABR,BR},
the latter demanding a beam of atoms to cross a high-Q cavity. The selective
Lindbladians are engineered by assuming the atomic level configuration in Fig.
1(a), the emission or absorption process following from the atoms prepared in
the ground $\left\vert g\right\rangle $ or excited $\left\vert e\right\rangle
$ state, respectively. The (non-selective) Lindbladian accounting for photon
absorption is engineered from the level configuration in Fig. 1(b). As shown
in Fig. 1(a), the cavity mode ($\omega$) is used to promote a Raman-type
transition $g\leftrightarrow e$, helped by two laser beams, $\omega_{1}$ and
$\omega_{2}$, out of tune with transitions $g\leftrightarrow i$ and
$e\leftrightarrow i$, respectively. In the configuration in Fig 1(b) ---which
follows from that in Fig. 1(a) by taking advantage of the Stark effect and
switching off the laser beams--- the cavity mode is now used to couple
resonantly the shifted levels $\left\vert g\right\rangle $ and $\left\vert
i\right\rangle $. When the configuration is that in Fig 1(a), the atoms are
randomly prepared in the states $\left\vert g\right\rangle $ and $\left\vert
e\right\rangle $, and when that in Fig. 1(b), the atoms are prepared at the
auxiliary level $\left\vert i\right\rangle $. Starting with the engineering of
the selective JC interactions arising from the diagram in Fig. 1(a), we write
the Hamiltonian%

\begin{figure}
[ptb]
\begin{center}
\includegraphics[
height=8.0cm,
width=13.0cm
]%
{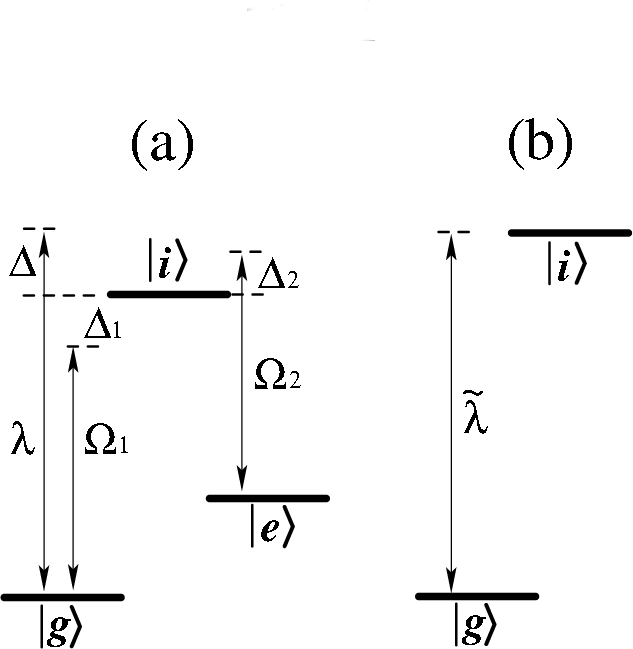}%
\caption{Atomic level configurations to engineer (a) selective and (b)
non-selective Hamiltonians.}%
\end{center}
\end{figure}
%

\begin{equation}
H=\lambda\sigma_{ig}a\operatorname*{e}\nolimits^{-i\Delta t}+\Omega_{1}%
\sigma_{ig}\operatorname*{e}\nolimits^{i\Delta_{1}t}+\Omega_{2}\sigma
_{ie}\operatorname*{e}\nolimits^{-i\Delta_{2}t}+H.c., \label{1}%
\end{equation}
where $\sigma_{rs}=\left\vert r\right\rangle \left\langle s\right\vert $, $r$
and $s$ labelling the atomic states involved, and we define $\Delta
=\omega-\omega_{ig}$, $\Delta_{1}=\omega_{ig}-\omega_{1}$, and $\Delta
_{2}=\omega_{2}-\omega_{ie}$, with $\omega_{i\ell}=\omega_{i}-\omega_{\ell}$
($\ell=g,e$). It is straightforward to verify that the conditions
$\lambda\sqrt{n+1}\ll\Delta$ and $\Omega_{j}\ll$ $\Delta_{j}$ ($j=1,2$) lead
to the effective interaction (\cite{James})
\begin{align}
H_{eff}  &  =\left(  \xi a^{\dag}a-\varpi_{g}\right)  \sigma_{gg}+\varpi
_{e}\sigma_{ee}\nonumber\\
&  +\left(  \zeta a^{\dagger}\operatorname*{e}\nolimits^{i\delta t}\sigma
_{ge}+H.c.\right)  , \label{2}%
\end{align}
where $\varpi_{g}=\left\vert \Omega_{1}\right\vert ^{2}/\Delta_{1}$ and
$\varpi_{e}=\left\vert \Omega_{2}\right\vert ^{2}/\Delta_{2}$ stand for
frequency level shifts due to the action of the classical fields, whereas the
strengths $\xi=\left\vert \lambda\right\vert ^{2}/\Delta$ and $\zeta
=\lambda^{\ast}\Omega_{2}\left(  \Delta^{-1}+\Delta_{2}^{-1}\right)  /2$ stand
respectively for off- and on-resonant atom-field couplings to be used to
engineer the required selective interactions; finally,\textbf{ }$\delta
=\Delta-\Delta_{2}$\textbf{ }refers to a convenient detuning to be specified
in the following lines. To get selectivity, we first perform the unitary
transformation $U=\exp\left\{  -i\left[  \left(  \xi a^{\dag}a+\varpi
_{g}\right)  \sigma_{gg}+\varpi_{e}\sigma_{ee}\right]  t\right\}  $, which
takes $H_{eff}$ into the form%
\begin{equation}
V_{eff}=%
{\textstyle\sum\nolimits_{n=1}^{\infty}}
\zeta_{n}\left\vert n+1\right\rangle \left\langle n\right\vert \sigma
_{ge}\operatorname*{e}\nolimits^{i\phi_{n}t}+H.c.\text{.} \label{3}%
\end{equation}
with $\zeta_{n}=\sqrt{n+1}\zeta$ and $\phi_{n}=\left(  n+1\right)  \xi
+\delta-\varpi_{g}-\varpi_{e}$. Next, under the strongly off-resonant regime
$\xi\gg\sqrt{k+2}\left\vert \zeta\right\vert $ and the condition%

\begin{equation}
\phi_{k}=0\text{,} \label{4}%
\end{equation}
which is easily satisfied by imposing $\left(  m+1\right)  \xi$ $=$
$\varpi_{g}$ $\gg\delta=\varpi_{e}$, such that $\left\vert \Omega
_{1}\right\vert =\sqrt{\left(  m+1\right)  \Delta_{1}/\Delta}\left\vert
\lambda\right\vert \gg\sqrt{\Delta_{1}/\Delta_{2}}\left\vert \Omega
_{2}\right\vert $, we readily eliminate, via RWA, all the terms proportional
to $\zeta_{n}=\sqrt{n+1}\zeta$ summed in $V_{eff}$, except when $n=k$,
bringing about the selective interaction%
\begin{equation}
\mathcal{H}_{1}=\left(  \zeta_{k}\left\vert k+1\right\rangle \left\langle
k\right\vert \sigma_{ge}+H.c.\right)  , \label{5}%
\end{equation}
producing the desired selective $g\leftrightarrow e$ transition within the
Fock subspace $\left\{  \left\vert k\right\rangle ,\left\vert k+1\right\rangle
\right\}  $. The excellent agreement between this effective selective
interaction and the full Hamiltonian (\ref{1}) has been analyzed in detail in
Ref. \cite{WilsonJOPB}. Regarding the Hamiltonian associated with the diagram
in Fig. 1(b), we readily see that switching off the laser field and tuning the
cavity mode to resonance with the atomic transition $g\leftrightarrow i$
results in:
\begin{equation}
\mathcal{H}_{2}=\tilde{\lambda}\sigma_{ig}a+H.c.. \label{6}%
\end{equation}

Next, following to the reasoning in Ref. \cite{ABR,BR} for atomic reservoir
engineering, we assume a weak-coupling regime for the interaction parameter
associated with $\mathcal{H}_{2}$, i.e., $\tilde{\lambda}\tau\ll1$, $\tau$
being the time during which each atom crosses the cavity. However, it is
easily verified that the Lindblad structure of the superoperator emerging from
$\mathcal{H}_{1}$ does not rely on the weak-coupling regime $\zeta_{k}\tau
\ll1$, owing to the selective nature of this interaction. When the atoms are
randomly prepared in the ground, excited, and auxiliary states: $p_{g}%
\sigma_{gg}+p_{e}\sigma_{ee}+p_{i}\sigma_{ii}$, with the laser detuning
$\Delta_{L}$ adjusted to produce $k=m$ and $l$, respectively, we obtain the
master equation \cite{ABR}%

\begin{align}
\frac{d\rho}{dt}  &  {\small =}\frac{\gamma_{m}}{2}\left(  2a_{m}\rho
a_{m}^{\dag}-\rho a_{m}^{\dag}a_{m}-a_{m}^{\dag}a_{m}\rho\right) \nonumber\\
&  {\small +}\frac{\gamma_{{\footnotesize l}}}{2}\left(  2a_{l}^{\dag}\rho
a_{l}-\rho a_{l}a_{l}^{\dag}-a_{l}a_{l}^{\dag}\rho\right) \nonumber\\
&  {\small +}\frac{\tilde{\gamma}}{2}\left(  2a^{\dag}\rho a-\rho aa^{\dag
}-aa^{\dag}\rho\right)  +\mathcal{L}\rho, \label{7}%
\end{align}
with the effective rates $\gamma_{m}=r^{g}\left(  \zeta_{m}\tau\right)  ^{2}$,
$\gamma_{{\footnotesize l}}=r^{e}\left(  \zeta_{l}\tau\right)  ^{2}$, and
$\tilde{\gamma}=r^{i}\left(  \tilde{\lambda}\tau\right)  ^{2}=\varepsilon
\gamma$ ($\varepsilon<1$ to achieve a steady equilibrium state), where $r^{g}%
$, $r^{e}$, and $r^{i}$ are the atomic arrival rates proportional to the
probabilities $p_{g},$ $p_{e}$, and $p_{i}$, respectively. The last term in
Eq. (\ref{7}), $\mathcal{L}\rho$, stands for the inevitable Liouvillian
operator describing the lossy cavity ($\omega$) of damping rate $\gamma$ and
temperature $T=\hbar\omega/k_{B}\ln\left[  \left(  1+\bar{n}\right)  /\bar
{n}\right]  $, $k_{B}$ being the Boltzmann constant, irrespective of the
passage of the atoms:
\begin{align}
\mathcal{L}\rho &  =\frac{\gamma}{2}\left(  1+\bar{n}\right)  \left(  2a\rho
a^{\dag}-\rho a^{\dag}a-a^{\dag}a\rho\right) \nonumber\\
&  {\small +}\frac{\gamma}{2}\bar{n}\left(  2a^{\dag}\rho a-\rho aa^{\dag
}-aa^{\dag}\rho\right)  . \label{8}%
\end{align}
From the equation of motion for the number state population, $\rho
_{nn}=\left\langle n\right\vert \rho\left\vert n\right\rangle $, derived from
Eqs. (\ref{7}) and (\ref{8}), we obtain the steady state solution (assuming
$l+1<m$):%

\begin{equation}
\rho_{nn}=\left\{
\begin{array}
[c]{cc}%
R_{n}\rho_{0} & n\leq l\\
R_{n}\mathcal{A}_{l}\rho_{0} & l+1\leq n\leq m\\
R_{n}\mathcal{B}_{l,m}\rho_{0} & n\geq m+1
\end{array}
\right.  , \label{9}%
\end{equation}
where $R_{n}=\left[  \left(  \varepsilon+\bar{n}\right)  /(1+\bar{n})\right]
^{n}$ and
\begin{subequations}
\label{10}%
\begin{align}
\mathcal{A}_{l}  &  =\frac{\gamma_{{\footnotesize l}}+(l+1)\left(
\varepsilon+\bar{n}\right)  \gamma}{(l+1)\left(  \varepsilon+\bar{n}\right)
\gamma},\label{10a}\\
\mathcal{B}_{l,m}  &  =\frac{(m+1)(1+\bar{n})\gamma}{\gamma_{{\footnotesize m}%
}+(m+1)\left(  1+\bar{n}\right)  \gamma}\mathcal{A}_{l},\label{10b}\\
\rho_{0}  &  =\frac{\left(  1-\varepsilon\right)  /(1+\bar{n})}{1-R_{l+1}%
+\mathcal{A}_{l}\left(  R_{l+1}-R_{m+1}\right)  +\mathcal{B}_{l,m}R_{m+1}}.
\label{10c}%
\end{align}

From Eqs. (\ref{9}) and (\ref{10}) we clearly see that the distribution
function $\rho_{nn}$ can be manipulated by an appropriate choice of the
engineered parameters $\gamma_{{\footnotesize l}}$, $\gamma_{{\footnotesize m}%
}$, and $\tilde{\gamma}$. To estimate the range of validity of these
parameters in a microwave cavity QED experiment, we start by choosing
$\Delta=\Delta_{1}=$ $(1+10^{-2})\times\Delta_{2}=10\sqrt{k+1}\left\vert
\lambda\right\vert $, such that $\left\vert \Omega_{1}\right\vert
=10\times\left\vert \Omega_{2}\right\vert =\sqrt{k+1}\left\vert \lambda
\right\vert $, $\zeta_{k}=10^{-2}\sqrt{k+1}\left\vert \lambda\right\vert $,
$\tau=10^{2}/\sqrt{k+1}\left\vert \lambda\right\vert $ (so that $\zeta_{k}%
\tau=1$ \cite{OBS}). Therefore, assuming $m,l\sim10$ with typical $\lambda
\sim5\times10^{5}$Hz and $\gamma\sim7.5$Hz, and imposing $\tilde{\gamma}\sim
r^{i}\times10^{-2}=\varepsilon\gamma$, it follows that $\gamma_{k}$ ranges up
to the order of $2\times10^{3}\gamma$, with $r^{i}=10^{3}\varepsilon$ (i.e.,
$p_{i}=10^{-2}\varepsilon$) and $r^{g},r^{e}$ ranging up to around $10^{4}$
($p_{g},p_{e}$ limited to $1-2p_{i}$).
\begin{figure}

\includegraphics[
height=10.0cm,
width=15.0cm
]%
{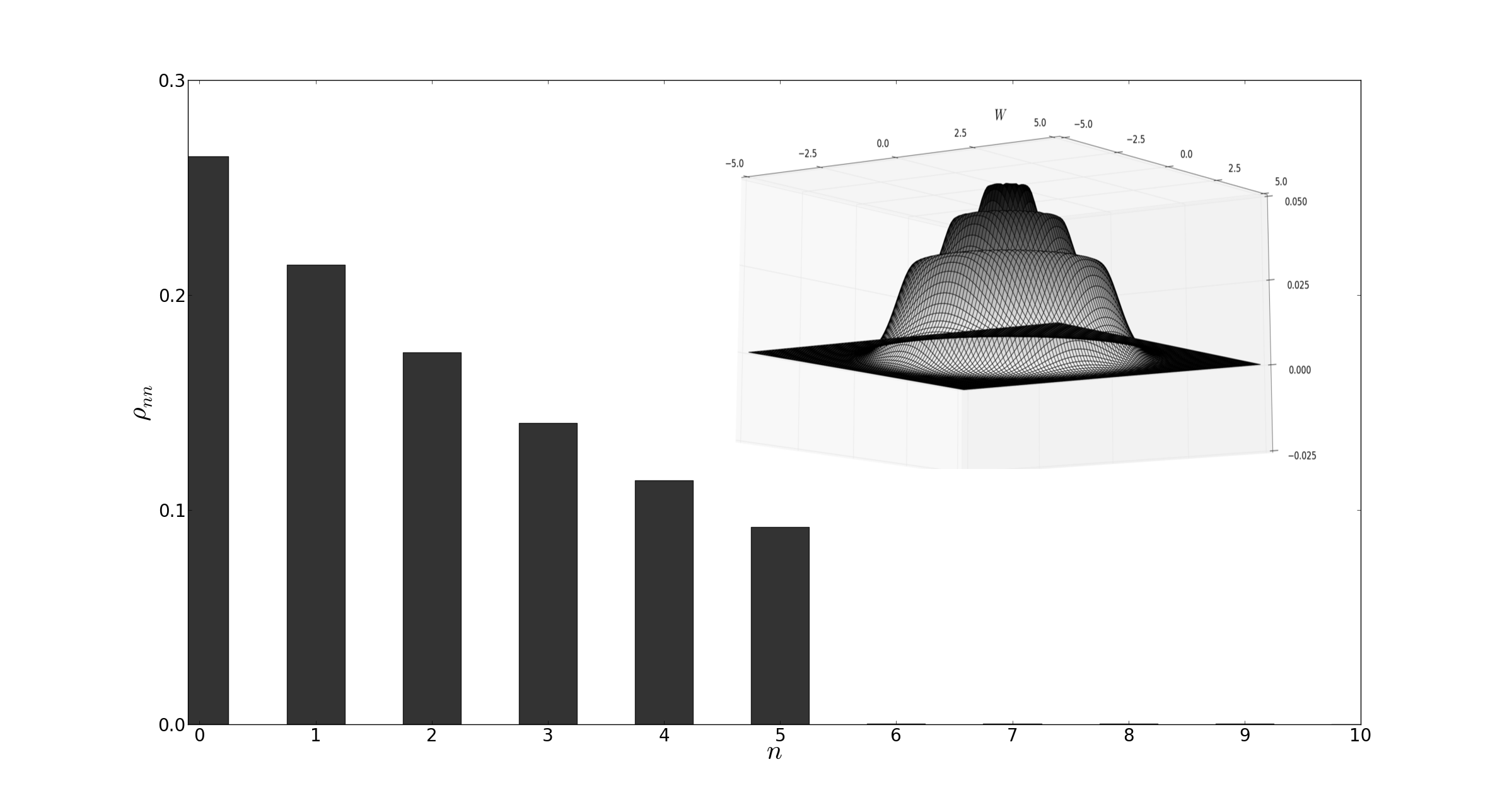}%
\caption{Truncated thermal distribution from the Fock state $m+1=6$. In the
inset we present the Wigner function of the truncated distribution.}%

\end{figure}

\begin{figure}
\includegraphics[
height=10.0cm,
width=15.0cm
]%
{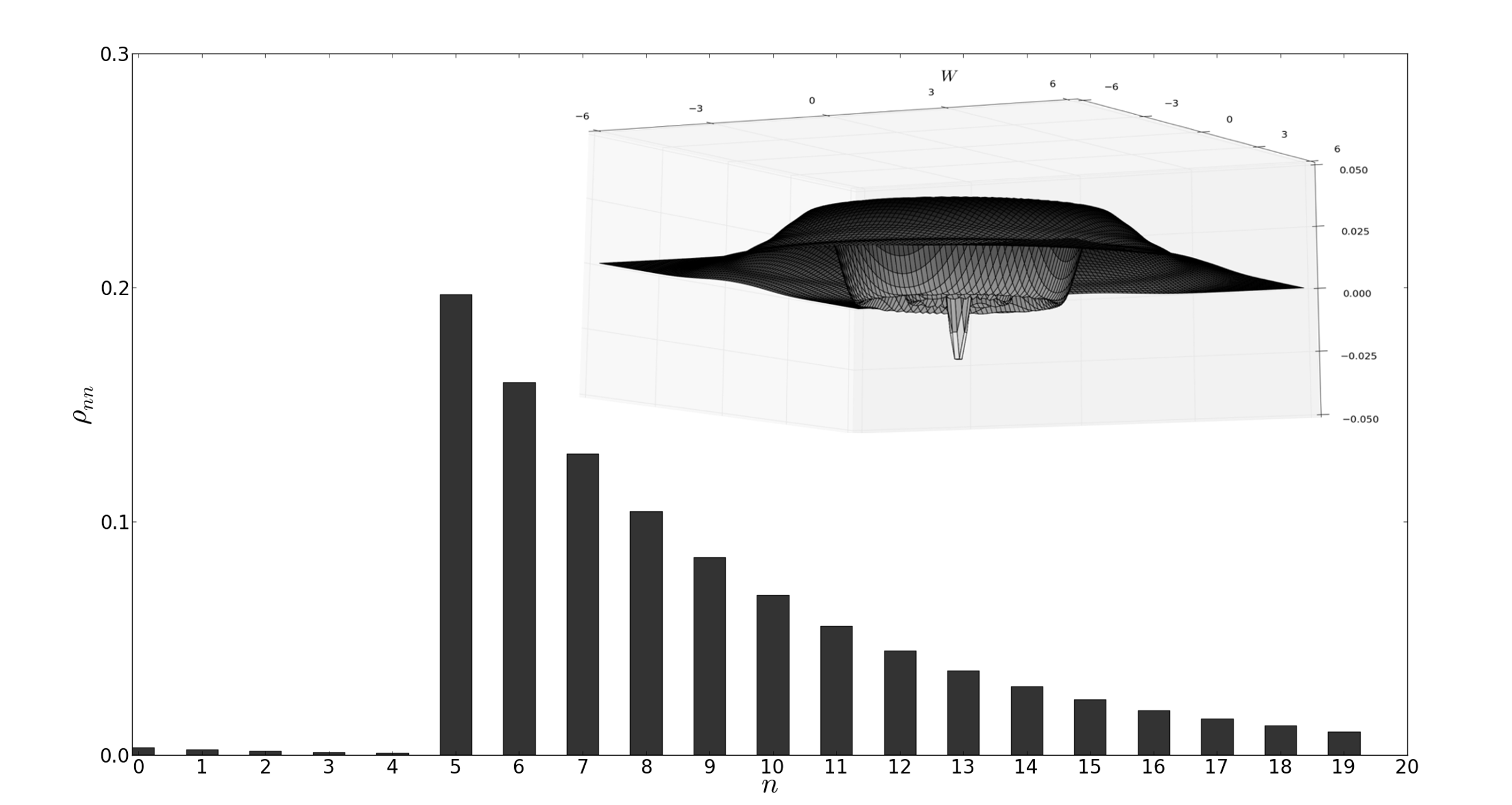}%

\caption{ Amplification of the thermal state distributions, from the Fock
state  $l+1=5$, with the corresponding Wigner distribution in the inset.}%

\includegraphics[
height=10.0cm,
width=15.0cm
]%
{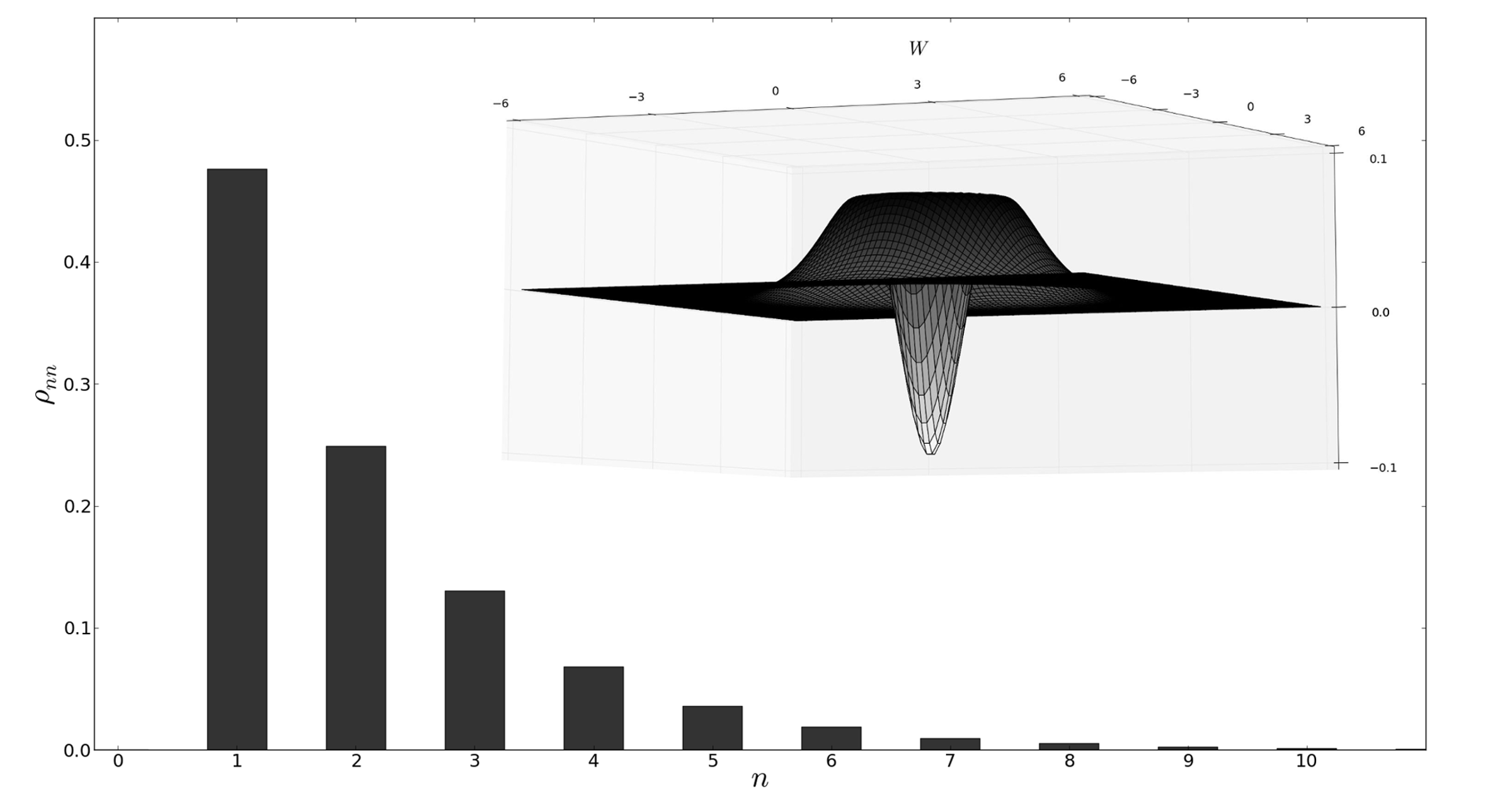}%

\caption{ Amplification of the thermal state distributions, from the Fock
state  $l+1=1$ , with the corresponding Wigner distribution in the inset.}%
\end{figure}

\begin{figure}
\begin{center}

\includegraphics[
height=10.0cm,
width=15.0cm
]%
{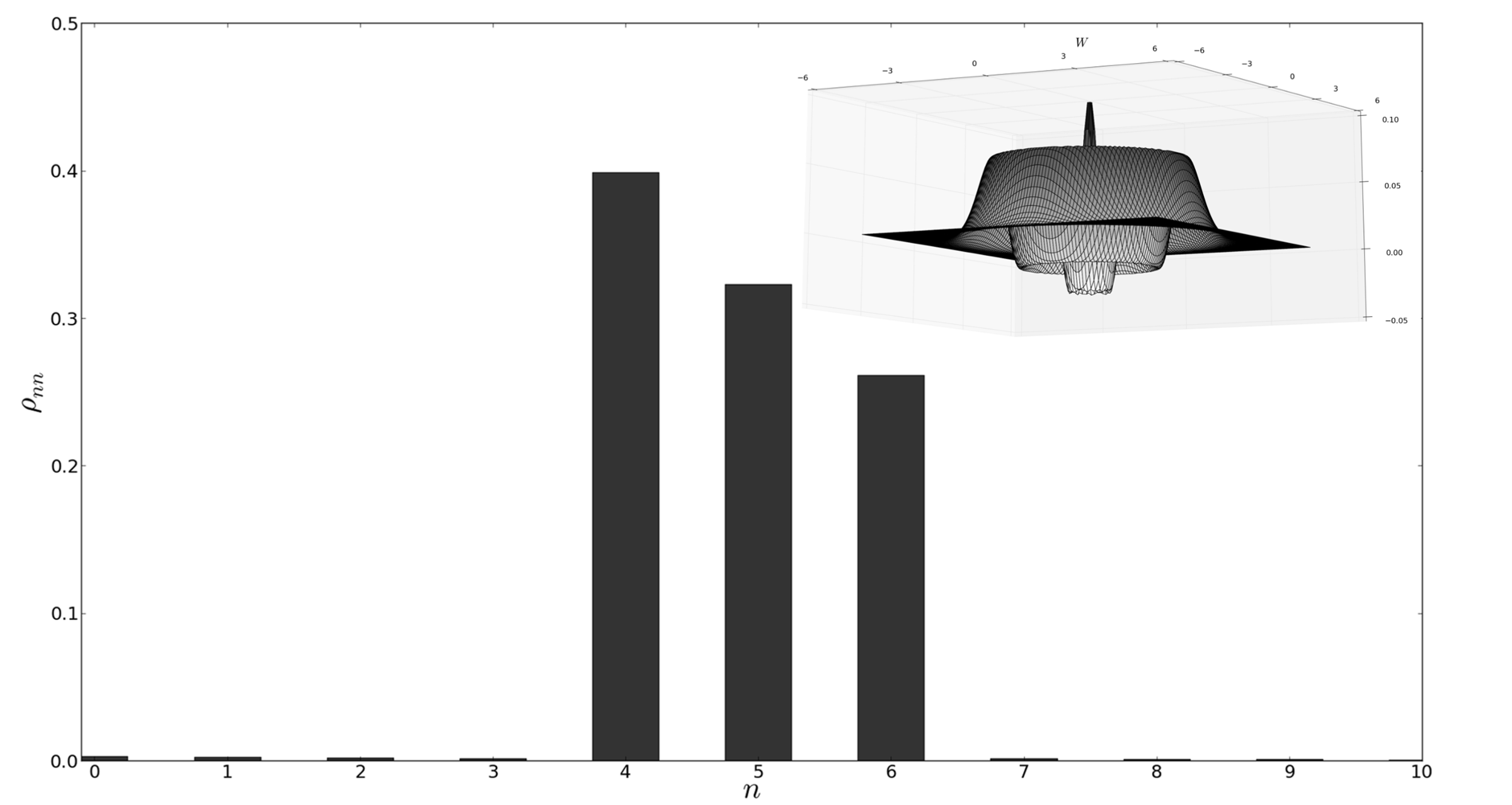}%
\caption{Sliced thermal distribution, from $l+1=4$ to $m=6$, with the
corresponding Wigner distribution in the inset.}%
\end{center}

\end{figure}
\begin{figure}
\begin{center}
\includegraphics[
height=10.0cm,
width=15.0cm
]%
{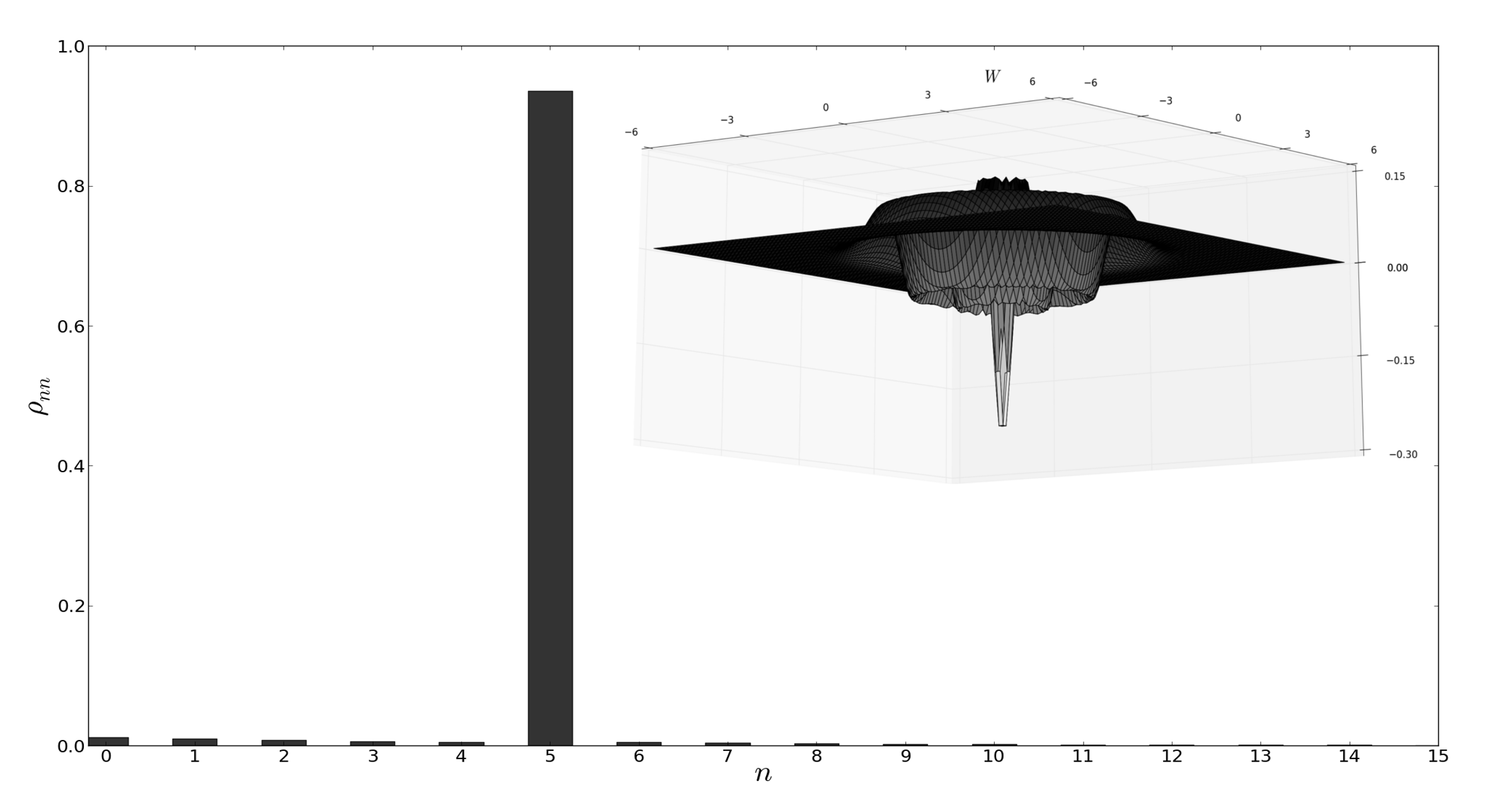}%

\caption{Steady Fock state  $\left\vert 5\right\rangle $ , with the corresponding Wigner distribution in the inset.}%

\includegraphics[
height=10.0cm,
width=15.0cm
]%
{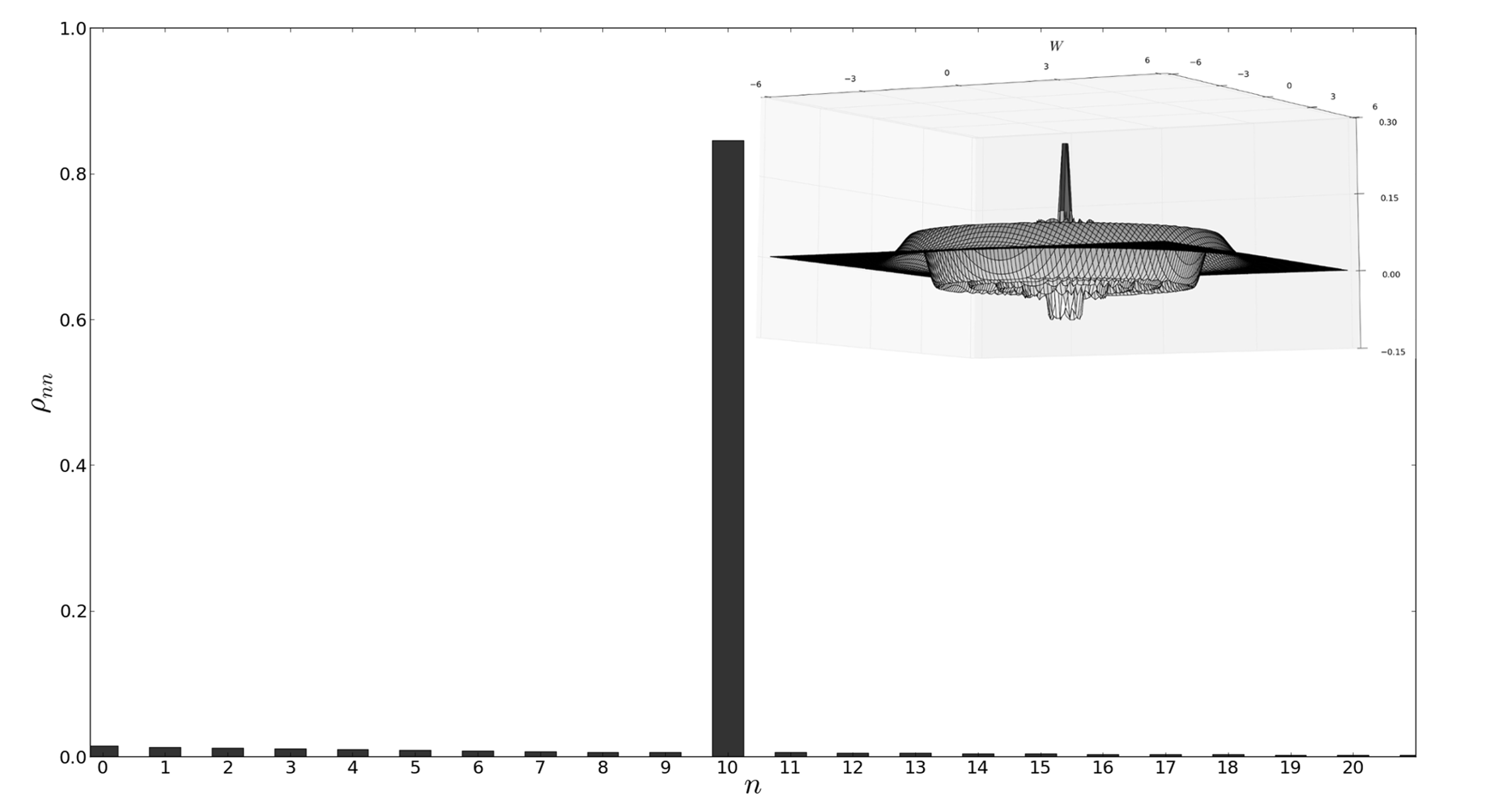}%

\caption{Steady Fock state  $\left\vert
10\right\rangle $, with the corresponding Wigner distribution in the inset.}%

\end{center}

\end{figure}
In order to generate steady Fock states: $i)$ we first observe that a
significantly large $\gamma_{{\footnotesize m}}$ (relative to $\gamma$), with
$\gamma_{{\footnotesize l}}=0$ (no atoms crossing the cavity in the excited
state), such that
\end{subequations}
\begin{equation}
R_{m+1}\mathcal{B}_{l,m}\ll R_{m}\mathcal{A}_{l}\text{,} \label{11}%
\end{equation}
entails the truncation of the equilibrium distribution $\rho_{nn}$, from the
population of state $m+1$, as illustrated in Fig. 2, where we have assumed
$m=5$, with $\bar{n}=0.05$ and $\varepsilon=0.8$, such that $\tilde{\gamma
}=0.8\gamma$ and $\gamma_{{\footnotesize m}}=10^{3}\gamma$ to satisfy Eq.
(\ref{11}). Although the analytical solution in Eqs. (\ref{9}) and (\ref{10})
was crucial to alert us to the possibility of manipulating the populations
$\rho_{nn}$ through the appropriate choice of the parameters involved, all the
simulations in this letter were based on Eq. (\ref{7}), running in QuTIP
\cite{QuTIP}. In the inset of Fig. 2, we present the Wigner function of the
truncated distribution, which takes no negative values, showing that it is a
purely classical state.

$ii)$ On the other hand, a significantly large damping rate $\gamma
_{{\footnotesize l}}$ (relative to $\gamma$), with $\gamma_{{\footnotesize m}%
}=0$ (no atoms crossing the cavity in the ground state), leads to the effect
opposite to that sketched in Fig. 2, enhancing the probabilities $\rho_{nn}$
for $n\geq$ $l+1$, instead of cutting them off. The amplification procedure
imposes the restriction%

\begin{equation}
R_{l+1}\mathcal{A}_{l}\gg R_{0}\text{.}\label{12}%
\end{equation}

\bigskip%

In Fig. 3, we illustrate the amplification of the equilibrium state, from
$l+1=5$, adopting the same parameters as in Fig. 2 but interchanging
$\gamma_{{\footnotesize m}}$ and $\gamma_{{\footnotesize l}}$, such that
$\gamma_{{\footnotesize l}}=10^{3}\gamma$. The Wigner function in the inset,
now exhibiting negative values, shows that this time we obtain a nonclassical
state of the radiation field. It is interesting to note, by the way, that by
simply eliminating the vacuum state from the thermal distribution, with $l=0$,
we automatically obtain a nonclassical state, as shown in Fig. 4, where we
use the same parameters as in Fig. 3, but with $\varepsilon=0.5$ (such that
$\tilde{\gamma}=0.5\gamma$), to prevent the excitation of large Fock states.
Therefore, the vacuum plays a major role in the intersection between
classicality and nonclassicality.

Another case arises when we put together $i)$ and $ii)$, so as to $iii)$ slice
the equilibrium distribution, from $l+1$ to $m$. This is done by ensuring the
conditions leading simultaneously to Eqs. (\ref{11}) and (\ref{12}). Fig. 5
shows a sliced steady distribution ranging from $l+1=4$ to $m=6$, again
assuming the same parameters as above, but $\gamma_{{\footnotesize m}}%
=\gamma_{{\footnotesize l}}=10^{3}\gamma$. The Wigner function in the inset
now exhibits a large region with negative values, strengthening the
nonclassical character of the generated state.
Finally, we come to the main point of our work: the choice $m=l+1$ allows us
to join the subspaces $\left\{  l,l+1\right\}  $ and $\left\{  m,m+1\right\}
$ alongside one another, so that both sare the same state $m$. This leads to
steady Fock states under the same conditions established in $iii)$, as seen in
Fig. 6, which illustrates the state $m=5$, prepared with exactly the same
parameters as in Fig. 5. The fidelity \cite{Livro} of the prepared Fock state
is around $0.97$, as confirmed by the Wigner distribution in the inset, which
exhibits the peculiar feature of the Fock state. We end up with the Fock state
$m=l+1=10$ in Fig. 7, reached with the same parameters in Fig. 4, except
$\varepsilon=0.95$, with the fidelity dropped to around $0.88$.

We have thus proposed a scheme to manipulate the steady thermal distribution
in such a way as to produce steady Fock states of the radiation field. Our
proposal relies on the engineering of selective JC Hamiltonians, which thus
generate equally selective Lindblad superoperators that enable us to
manipulate the equilibrium thermal distribution, slicing it so as to prepare
steady Fock states. Our technique can be implemented in other contexts of
atom-field interaction, such as trapped ions and circuit QED, where the beam
of atoms simulating the reservoir can be achieved by a pulsed classical field.
In the former case, the classical field is used to couple the vibrational
field intermittently with the internal ionic states, while in the latter case,
it is used to bring a cooper-pair box into resonance with the mode of a
superconducting strip. Apart from the preparation of Fock states, other
applications within Hamiltonian, reservoir, and state engineering may arise
from the present protocol, as for example the generation of entangled steady
state in a network of quantum oscillators \cite{G}.

The authors acknowledge financial support from PRP/USP within the Research
Support Center Initiative (NAP Q-NANO) and FAPESP, CNPQ and CAPES, Brazilian agencies.


\begin{thebibliography}{9999}
\bibitem {OBS}We performed (following the reasoning in Refs. \cite{ABR,BR}) a
numerical simulation of the passage of atoms through the cavity, assuming
$\tau=10^{-4}$s in conformity with $\zeta_{k}\tau=1$, to confirm that the
cavity mode evolves in excellent agreement with the coarse-graining master
equation (\ref{7}), and that higher value of $\tau$ start to compromise the
accuracy of this master equation.
\end{thebibliography}
\end{document}